\documentclass[12pt]{article}

\usepackage{graphicx}
\usepackage{amssymb,amsmath,slashed}
\usepackage{wrapfig}

\def\pp{{\mbox{\boldmath$p$}}}
\def\thetap{\theta_{\mbox{\bfseries\itshape\scriptsize p}}}
\def\ep{E_{\mbox{\bfseries\itshape\scriptsize p}}}
\def\q{{\mbox{\boldmath$q$}}}

\def\cm{^{\rm \mbox{\tiny CM}}}

\def\p'{{\mbox{\boldmath$p'$}}}
\def\k{{\mbox{\boldmath$k$}}}
\def\ek{E_{\mbox{\bfseries\itshape\scriptsize k}}}
\def\l{{\mbox{\boldmath$l$}}}

\def\vz{{\mbox{\boldmath$0$}}}
\def\thetaps{\theta_{\mbox{\bfseries\itshape\scriptsize {p$^*$}}}}

\def\boa{B$_{I}$}
\def\bob{B$_{II}$}
\def\boc{B$_{III}$}
\def\bod{B$_{IV}$}
\def\boe{B$_{V}$}

\def\ee{$E_e$}
\def\eepr{$E_e^{\prime}$}
\def\thepr{$\theta$}

\def\thn{$\theta_n$}

\def\thp{$\theta_p$}
\def\thqe{$\theta_{qe}$}
\def\thpe{$\theta_{pe}$}
\def\sqs{$\sqrt{s}$}
\def\sqsmm{$\sqrt{s}-2m$}
\begin{document}

\begin{center}
{\bf\large Final state interaction effects in exclusive\\
electrodisintegration of the deuteron}
\end{center}

\vskip 10mm

\noindent
{\bf S. G. Bondarenko}\footnote{Speaker.
XXI International Baldin Seminar on High Energy Physics Problems.
                 September 10-15, 2012.
                 JINR, Dubna, Russia}\\
Joint Institute for Nuclear Research, Dubna, Russia\\
E-mail: {\it bondarenko@jinr.ru}\\[5mm]
{\bf V. V. Burov}\\
Joint Institute for Nuclear Research, Dubna, Russia\\
E-mail: {\it burov@theor.jinr.ru}\\[5mm]
{\bf E. P. Rogochaya}\\
Joint Institute for Nuclear Research, Dubna, Russia\\
E-mail: {\it rogochaya@theor.jinr.ru}

\vskip 10mm

\begin{minipage}{160mm}
The exclusive electrodisintegration of the deuteron
is considered within the Bethe-Salpeter approach with a separable interaction
kernel. The relativistic kernel of nucleon-nucleon interaction is obtained
considering the phase shifts in the elastic neutron-proton scattering and
properties of the deuteron.
The differential cross section is calculated within the impulse approximation under
several kinematic conditions of the Bonn experiment. Final state
interactions between the outgoing nucleons are taken into account.
Partial-wave states of the neutron-proton pair with
total angular momentum $J=0,1$ are considered.
%}
\end{minipage}

\section{Introduction}
The exclusive electrodisintegration of the deuteron is a useful instrument
which makes it possible to investigate the electromagnetic
structure not only of the bound state - deuteron but also the
scattering states of the neutron-proton ($np$) system. Many approaches
have been elaborated to describe this reaction for the last 40 years
\cite{Forest:1983vc,Arenhovel:1982rx,Shebeko,Gakh:2004zq,Jeschonnek:2008zg}.
The simplest of them considered the electrodisintegration within a
nonrelativistic model of the nucleon-nucleon (NN) interaction and
outgoing nucleons were supposed to be free \cite{Forest:1983vc}
(the plane-wave approximation - PWA). Those approaches were in a
good agreement with experimental data at low energies. However,
further investigations have shown that the final state interaction
(FSI) between outgoing nucleons, two-body currents and other
effects should be taken into account to obtain a reasonable
agreement with existing experimental data at higher energies. Most
of these effects have been considered within nonrelativistic
models \cite{Arenhovel:1982rx,Shebeko}. In relativistic models,
FSI effects could be calculated within quasipotential approaches
using the on-mass-shell nucleon-nucleon $T$ matrix
\cite{Gakh:2004zq,Jeschonnek:2008zg}.

One of the fundamental approaches used to describe of the $np$ system
is based on the Bethe-Salpeter (BS) equation
\cite{Salpeter:1951sz}. The separable ansatz \cite{Yamaguchi:1954mp}
for the NN interaction kernel was successfully applied to solve this equation.
However, the approach was not applicable for the most
of high-energy NN processes for a while since calculated
expressions contained nonintegrable singularities. The problem was solved
in \cite{Bondarenko:2008fp,Bondarenko:2010qv,Bondarenko:2008mm}
where kernels of special type were proposed.
Using them FSI can be taken into account (in particular,
considering the electrodisintegration) in a wide range of energy.

In the present paper, the electrodisintegration cross section is
calculated under different kinematic
conditions of Bonn experiments~\cite{Breuker:1985sz,Boden:1992jz}.
The rank-six NN interaction potential MY6 \cite{Bondarenko:2010qv}
is used to describe scattered states $^3S_1$-$^3D_1$ and the deuteron.
The uncoupled partial-wave states with total angular momentum
$J=0,1$ ($^1S_0$, $^1P_1$, $^3P_0$, $^3P_1$) are described by
multirank separable potentials \cite{Bondarenko:2008mm}.

The paper is organized as follows. In Sec.\ref{sec:2}, the
exclusive three-differential cross section of the $d(e,e^\prime p)n$
reaction is calculated within the relativistic impulse approximation.
The used BS formalism is presented in Sec.\ref{sec:3}. The details
of calculations are considered in Sec.\ref{sec:4}. Then the
obtained relativistic results are compared with
experimental data of the Bonn experiments~\cite{Breuker:1985sz,Boden:1992jz}
in Sec.\ref{sec:5}.
\section{Cross section}\label{sec:2}
When all particles are unpolarized the exclusive
electrodisintegration of the deuteron $d(e,e^\prime p)n$ can be
described by the differential cross section in the deuteron rest
frame - laboratory system (LS), which has the following form:
\begin{eqnarray}
&&\frac{d^3\sigma}{dE_e'd\Omega_e'd\Omega_p}=
\frac{\sigma_\textmd{Mott}}{8M_d(2\pi)^3}{\frac{\pp_p^2{\sqrt s}}{\sqrt{1+\eta}|\pp_p|-E_p\sqrt{\eta}
\cos\theta_p}}
\label{3cross_0}\\
&&\times\left[ l^0_{00}W_{00}+l^0_{++}(W_{++}+W_{--})+ 2 l^0_{+-}\cos2\phi~{\rm Re}
W_{+-}
%\nonumber\\
%&&\hspace*{5mm}
- 2 l^0_{+-}\sin2\phi~{\rm Im} W_{+-}
\right.
\nonumber\\
&&\hspace*{5mm}
\left.
-2 l^0_{0+}\cos\phi~{\rm Re} (W_{0+}-W_{0-})
-2 l^0_{0+}\sin\phi~{\rm Im} (W_{0+}+W_{0-})\right]. \nonumber
\end{eqnarray}
where $\sigma_{\rm
Mott}=(\alpha\cos\frac{\theta}{2}/2E_e\sin^2\frac{\theta}{2})^2$
is the Mott cross section, $\alpha={e^2}/{(4\pi)}$ is the fine
structure constant; $M_d$ is the mass of the deuteron;
$q=p_e-p_e'=(\omega,\q)$ is the momentum transfer; $p_e=(E_e,\l)$
and $p_e'=(E_e',\l')$ are initial and final electron momenta,
respectively; $\Omega_e'$ is the outgoing electron solid angle;
$\theta$ is the electron scattering angle. The outgoing proton is
described by momentum $\pp_p$ ($E_p=\sqrt{\pp_p^2+m^2}$, $m$ is
the mass of the nucleon) and solid angle
$\Omega_p=(\theta_p,\phi)$ where $\theta_p$ is the zenithal angle
between the $\pp_p$ and $\q$ momenta and $\phi$ is the azimuthal
angle between the ({\bf ee$^\prime$}) and ({\bf qp}) planes.
Factor $\eta=\q^2/s$ can be calculated through the $np$ pair total
momentum $P$ squared:
\begin{eqnarray}
s=P^2=(p_p+p_n)^2
%=(M_d+\omega)^2-\q^2
=M_d^2+2M_d\omega+q^2,
\label{pair_s}
\end{eqnarray}
defined by the sum of the proton $p_p$ and neutron $p_n$ momenta.
The photon density matrix elements have the following form:
\begin{eqnarray}
%&&
l_{00}^0=\frac{Q^2}{\q^2},\quad l_{0+}^0=\frac{Q}{|\q|\sqrt
2}\sqrt{\frac{Q^2}{\q^2}+\tan^2 \frac{\theta}{2}},
%\nonumber\\
%&&
l_{++}^0=\tan^2\frac{\theta}{2}+\frac{Q^2}{2\q^2},\quad
l_{+-}^0=-\frac{Q^2}{2\q^2},
\end{eqnarray}
where $Q^2=-q^2$ is introduced for convenience. The hadron density
matrix elements have the following form:
\begin{eqnarray}
W_{\lambda\lambda'}=W_{\mu\nu}\varepsilon_\lambda^\mu\varepsilon_{\lambda'}^\nu,
\end{eqnarray}
where $\lambda$, $\lambda'$ are photon helicity components
\cite{Dmitrasinovic:1989bf}, can be calculated using the photon
polarization vectors $\varepsilon$ and Cartesian components of
hadron tensor
\begin{eqnarray}
W_{\mu\nu}=\frac{1}{3}\sum_{s_ds_ns_p}\left|<np:SM_S|j_\mu|d:1M>\right|^2,
\label{ht}
\end{eqnarray}
where $S$ is the spin of the $np$ pair and $M_S$ is its
projection; $s_d$, $s_n$ and $s_p$ are deuteron, neutron and
proton momentum projections, respectively. The matrix element
$<np:SM_S|j_\mu|d:1M>$ can be written using the
Mandelstam technique \cite{Mandelstam:1955sd} as follows:
\begin{eqnarray}
<np:SM_S|j_\mu|d:1M>=i\sum_{n=1,2}\int\frac{d^4p\cm}{(2\pi)^4}
%\times
\label{cur_fsi}
%\\
%&&
{\rm Sp}\left\{\Lambda({\cal L
}^{-1})\bar\psi_{SM_S}(p\cm,{p^*}\cm;P\cm) \Lambda({\cal L
})
\right.
\times
%\nonumber
\\
\hskip 30mm
\Gamma_\mu^{(n)}(q)
S^{(n)}\left(\frac{K_{(0)}}{2}-(-1)^np-\frac q2\right)
\left.\Gamma^M
\left(p+(-1)^n\frac{q}{2};K_{(0)}\right)\right\}\nonumber
\end{eqnarray}
within the relativistic impulse approximation in LS. The sum over
$n=1,2$ corresponds to the interaction of the virtual photon with
proton and with neutron in the deuteron, respectively. The
total $P\cm$ and relative ${p^*}\cm$ momenta of the outgoing
nucleons and the integration momentum $p\cm$ are considered in the
final $np$ pair rest frame - center-of-mass system (CMS), $p$
denotes the relative $np$ pair momentum in LS. To perform the
integration, momenta $p$, $q$ and deuteron total momentum
$K_{(0)}=(M_d,\vz)$ in LS are written in CMS using the Lorenz-boost
transformation ${\cal L}$ along the $\q$ direction. The $np$ pair
wave function $\psi_{SM_S}$ is transformed from CMS to LS applying the
corresponding boost operator $\Lambda$. A detailed description of
$\psi_{SM_S}$, the $n$th nucleon interaction vertex
$\Gamma_\mu^{(n)}$, the propagator of the $n$th nucleon $S^{(n)}$,
and the deuteron vertex function $\Gamma^M$ can be found in our
previous works \cite{Bondarenko:2002zz,Bondarenko:2005rz}.

\section{Separable kernel of NN interaction}\label{sec:3}
The outgoing $np$ pair is described by the $T$ matrix which can be
found solving the inhomogeneous Bethe-Salpeter equation
\cite{Salpeter:1951sz}:
\begin{eqnarray}
%&&
T(p^{\prime}, p; P) = V(p^{\prime}, p; P) \label{t_op}
% \\
%&&\hspace*{10mm}
+ \frac{i}{4\pi^3}\int d^4k\, V(p^{\prime}, k;
P)\, S_2(k; P)\, T(k, p; P),
%\nonumber
\end{eqnarray}
where $V$ is the NN interaction kernel, $S_2$ is the free
two-particle Green function:
\begin{eqnarray}
S_2^{-1}(k; P)=\bigl(\tfrac12\:{\slashed P}+{\slashed
k}-m\bigr)^{(1)} \bigl(\tfrac12\:{\slashed P}-{\slashed
k}-m\bigr)^{(2)},
\end{eqnarray}
and $p~(p')$ is the relative momentum of initial (final) nucleons,
$P$ is the total $np$ pair momentum.

To solve the BS equation (\ref{t_op}) partial-wave decomposition
\cite{Kubis:1972zp} for the $T$ matrix:
\begin{eqnarray}
T_{\alpha\beta,\gamma\delta}(p^{\prime},p; {P_{(0)}}) =
%&&
\sum_{JMab} t_{ab}(p_0',|\p'|;p_0,|\pp|; s)
% \times
\label{bse_spd}
%\\
%&&\hspace*{-7mm}
({\cal Y}_{aM}(-{\p'})U_C)_{\alpha\beta}\otimes
(U_C {\cal Y}^{\dag}_{bM}({\pp}))_{\delta\gamma}
\end{eqnarray}
is used. Here $P_{(0)}=(\sqrt s,\vz)$ is the $np$ pair total
momentum in CMS, $U_C=i\gamma^2\gamma^0$ is the charge conjugation
matrix. Indices $a,b$ correspond to the set $^{2S+1}L_J^\rho$ of
spin $S$, orbital $L$ and total $J$ angular momenta, $\rho=+$
defines a positive-energy partial-wave state, $\rho=-$ corresponds
to a negative-energy one. Greek letters
$\{\alpha,\beta,\gamma,\delta\}$ in (\ref{bse_spd}) are used to denote Dirac matrix
indices. The spin-angle functions:
\begin{eqnarray}
%&&
{\cal Y}_{JM:LS {\rho}}(\pp) U_C= \label{an_pa}
%\\
% &&
i^{L}\sum_{m_Lm_Sm_1m_2\rho_1\rho_2}C_{\frac12 \rho_1
\frac12 \rho_2}^{S_{\rho} {\rho}} C_{L m_L S m_S}^{JM} C_{\frac12
m_1 \frac12 m_2}^{Sm_S}
%\nonumber\\
%&&\hspace*{20mm}\times 
Y_{L{m_L}}(\hat\pp)
{U^{\rho_1}_{m_1}}^{(1)}(\pp) {{U^{\rho_2}_{m_2}}^{(2)}}^{T}(-\pp)
\nonumber
\end{eqnarray}
are constructed using free nucleon Dirac spinors $u,~v$. It
should be mentioned that only positive-energy states with $\rho=+$
are considered in this paper. Performing similar decomposition for
$V$, the BS equation for radial parts of the $T$ matrix and kernel $V$
is obtained:
\begin{eqnarray}
&&t_{ab}(p_0', |\p'|; p_0, |\pp|; s) = v_{ab}(p_0', |\p'|; p_0,
|\pp|; s)
\label{BS_decomp}\\
&&+ \frac{i}{4\pi^3}\sum_{cd}\int\limits_{-\infty}^{+\infty}\!
dk_0\int\limits_0^\infty\! \k^2 d|\k|\, v_{ac}(p_0', |\p'|; k_0,
|\k|; s)
%\nonumber\\
%&& \hspace*{20mm}\times
 S_{cd}(k_0,|\k|; s)\, t_{db}(k_0,|\k|;p_0,|\pp|;s).
\nonumber
\end{eqnarray}
To solve the resulting equation (\ref{BS_decomp}), a separable
ansatz \cite{Yamaguchi:1954mp} for the interaction kernel $v$ is
used:
\begin{eqnarray}
%&&
v_{ab}(p_0', |\p'|; p_0, |\pp|; s)=
\label{ansatz}
%\\
%&& \hspace*{10mm}
 \sum_{i,j=1}^N\lambda_{ij}(s) g_i^{[a]}(p_0',
|\p'|)g_j^{[b]}(p_0, |\pp|),
%\nonumber
\end{eqnarray}
where $N$ is a rank of a separable kernel, $g_i$ are model
functions, $\lambda_{ij}$ is a parameter matrix. Substituting $v$
(\ref{ansatz}) into BS equation (\ref{t_op}), we obtain the $t$
matrix in a similar separable form:
\begin{eqnarray}
%&&
t_{ab}(p_0', |\p'|; p_0, |\pp|; s)= 
\label{t_separ}
%\\
%&& \hspace*{10mm}
\sum_{i,j=1}^N\tau_{ij}(s)g_i^{[a]}(p_0', |\p'|)
g_j^{[b]}(p_0, |\pp|)
%\nonumber
\end{eqnarray}
where:
\begin{eqnarray}
\tau_{ij}(s)=1/(\lambda_{ij}^{-1}(s)+h_{ij}(s)),
\end{eqnarray}
and
\begin{eqnarray}
%&&
h_{ij}(s)= 
\label{hij}
%\\
%&& 
-\frac{i}{4\pi^3}\sum_{a}\int dk_0\int \k^2d|\k|
\frac{g_i^{[a]}(k_0,|\k|)g_j^{[a]}(k_0,|\k|)}{(\sqrt
s/2-\ek+i\epsilon)^2-k_0^2}
%\nonumber
\end{eqnarray}
are auxiliary functions, $\ek=\sqrt{\k^2+m^2}$. Thus, the problem
of solving the initial integral BS equation (\ref{t_op}) turns out
to finding functions $g_i$ and parameters $\lambda_{ij}$ of
separable representation (\ref{ansatz}). They can be obtained
from a description of observables in $np$ elastic scattering
\cite{Bondarenko:2010qv,Bondarenko:2008mm,Rupp:1989sg,Mathelitsch:1981mr,Bondarenko:2011hs}.

\section{Final state interaction}\label{sec:4}
To calculate FSI we need to consider the interacting
$np$ pair. The outgoing nucleons are described by the BS amplitude which can
be written as a sum of two terms:
\begin{eqnarray}
%&&
\psi_{SM_S}(p,p^*;P) = \psi^{(0)}_{SM_S}(p,p^*;P) \label{psi_all}
%\\
%&&
+\frac{i}{4\pi^3}
S_2(p;P)\int d^4k ~V(p,k;P)\psi_{SM_S}(k,p^*;P).
%\nonumber
\end{eqnarray}
The first term
\begin{eqnarray}
\psi_{SM_S}^{(0)} (p,p^*;P)=
(2\pi)^4\chi_{SM_S}(p;P)\delta(p-p^*)\label{wf_PWA}
\end{eqnarray}
is related to the outgoing pair of free nucleons,
$\chi_{SM_S}$ is a spinor function for two fermions. The second term in
(\ref{psi_all}) corresponds to the final state interaction of the
outgoing nucleons. If we use the following relation:
\begin{eqnarray}
%&&
\int d^4k~V(p,k;P)\psi_{SM_S}(k,p^*;P)=
%\\
%&&
\int d^4k~T(p,k;P)
\psi^{(0)}_{SM_S}(k,p^*;P)\nonumber
\end{eqnarray}
then it can be transformed into
\begin{eqnarray}
&&
\psi_{SM_S}^{(t)}(p,p^*;P)=
\label{psi_t}
%\\
%&& \hspace*{10mm} 
4\pi i S_2(p;P)T(p,p^*;P)\chi_{SM_S}(p^*;P),
%\nonumber
\end{eqnarray}
here $(t)$ means that this part of the $np$ pair wave function is
related to the $T$ matrix. Using the partial-wave decomposition
(\ref{bse_spd}) expression (\ref{psi_t}) can
be written as follows:
\begin{eqnarray}
%&&
\psi_{SM_S}^{(t)}(p,p^*;P)=  4\pi i
% \times
 \label{psi_t_decomposed}
%\\
%&&
\sum_{LmJMa} C_{LmSM_S}^{JM} Y_{Lm}^*(\hat\pp^*){\cal
Y}_{aM}(\pp) \phi_{a,J:LS+}( p_0,|\pp|; s), 
%\nonumber
\end{eqnarray}
where $p^*=(0,\pp^*)$ with $|\pp^*|=\sqrt{s/4-m^2}$ is the
relative momentum of on-mass-shell nucleons in CMS, $\hat\pp^*$
denotes the azimuthal angle $\thetaps$ between the $\pp^*$ and
$\q$ vectors and zenithal angle $\phi$. Since only
positive-energy partial-wave states are considered here the radial part
is:
\begin{eqnarray}
\phi_{a,J:LS+}(p_0,|\pp|; s)=\frac
{t_{a,J:LS+}(p_0,|\pp|;0,|\pp^*|; s)}{(\sqrt
s/2-\ep+i\epsilon)^2-p_0^2}.
\end{eqnarray}
According to definition (\ref{an_pa}) spin-angle functions ${\cal
Y}$ can be written as a product of Dirac matrices $\gamma$ in the
matrix representation \cite{Bondarenko:2002zz} as follows:
\begin{eqnarray}
%&&
{\cal Y}_{aM}(\pp)=
\label{sap_np}
%\\
%&&
\frac{1}{\sqrt{8\pi}}
\frac{1}{4\ep(\ep+m)}(m+{\slashed p}_1)(1+\gamma_0){\cal G}_{aM}
(m-{\slashed p}_2),
%\nonumber
\end{eqnarray}
matrices ${\cal G}_{aM}$ are shown in Table
\ref{tab:1}.
%\begin{eqnarray}
%\gamma_5=-i\gamma^0\gamma^1\gamma^2\gamma^3=\left(
%\begin{array}{cccc}
% 0 &  0 & -1 &  0 \\
% 0 &  0 &  0 & -1 \\
%-1 &  0 &  0 &  0 \\
% 0 & -1 &  0 &  0 \nonumber
%\end{array}
%\right).
%\end{eqnarray}.
Decomposition (\ref{psi_t_decomposed})
is considered in detail in \cite{Bondarenko:2004pn}.
\begin{table}[h]
\begin{center}
\begin{tabular}{cc}
\hline\hline
$a={\footnotesize \left\{^{2S+1}L_J^\rho\right\}^{{\phantom{1}}^{\phantom{1}}} }$ & ${\cal G}_{aM}$ \\
\hline
$~{^1S_0^+}^{{\phantom{1}}^{\phantom{1}}}$ & $-\gamma_5$ \\
$~^3S_1^+$ & ${\slashed \xi}_{M}$ \\
$~^1P_1^+$ & $\frac{\sqrt{3}}{|\pp|}(p_1\cdot\xi_{M})\gamma_5$ \\
$~^3P_0^+$ & $-\frac{1}{2|\pp|}({\slashed p_1}-{\slashed p_2})$ \\
$~^3P_1^+$ &
$-\sqrt{\frac{3}{2}}\frac{1}{|\pp|}\left[(p_1\cdot\xi_{M})-\frac{1}{2}
{\slashed \xi}_{M}({\slashed p_1}-{\slashed p_2})\right]\gamma_5$ \\
$~^3D_1^+$ & $\frac{1}{\sqrt{2}}\left[{\slashed
\xi}_{M}+\frac{3}{2}\frac{1}{\pp^2}(p_1\cdot\xi_{M})({\slashed p}_1-{\slashed
p}_2)\right]$
\\
\hline\hline
\end{tabular}
\caption{\small Spin-angular parts \protect${\cal G}_{aM}$ (\ref{sap_np}) for the
$np$ pair; $p_1=(\ep,\pp)$, $p_2=(\ep,-\pp)$ are on-mass-shell
momenta, \protect$\ep=\sqrt{\pp^2+m^2}$; $\gamma$ matrices are defined as in
\protect\cite{Bjorken}.} \label{tab:1}
\end{center}
\end{table}
Using definition (\ref{psi_all}) and substituting (\ref{wf_PWA}),
(\ref{psi_t_decomposed}) into (\ref{cur_fsi}), the final
expression for hadron current \\ $<np:SM_S|j_\mu|d:1M>$ can be
obtained. It consists of two parts. One of them:
\begin{eqnarray}
%&&
<np:SM_S|j_\mu|d:1M>^{(0)}=
\label{cur_0}
%\\
%&&
i\sum_{n=1,2}\left\{\Lambda({\cal
L}^{-1})\bar\chi_{SM_S}\left({p^*}\cm; P\cm\right)\Lambda({\cal L})
 \right.\times 
%\nonumber
\\
\Gamma_\mu^{(n)}(q)
%&&
S^{(n)}\left(\frac{K_{(0)}}{2}-(-1)^n
p^*-\frac{q}{2}\right)\left.\Gamma^{M}\left(p^*+(-1)^n\frac{q}{2};
K_{(0)}\right)\right\}\nonumber
\end{eqnarray}
corresponds to the electrodisintegration in PWA. Another one:
\begin{eqnarray}
<np:SM_S|j_\mu|d:1M>^{(t)}=
\hskip 103mm \label{cur_t}
\\
\frac{i}{4\pi^3}\sum_{n=1,2}\sum_{LmJM_JL'lm'}
C_{LmSM_S}^{JM_J} Y_{Lm}(\hat\pp^*)
\int\limits_{-\infty}^\infty dp_0\cm \int\limits_0^\infty
{(\pp\cm)}^2 d|\pp\cm| \int\limits_{-1}^1
d\cos\thetap\cm \int\limits_0^{2\pi}d\phi
\times \hskip 25mm
\nonumber\\
{\rm Sp}\left\{\Lambda({\cal L }^{-1})\bar{\cal
Y}_{JL'SM_J}(\pp\cm) \Lambda({\cal L })
\Gamma_\mu^{(n)}(q)\times \right.
 \left. S^{(n)}\left(\frac{K_{(0)}}{2}-(-1)^np-\frac
q2\right){\cal Y}_{1lSm'}\left(\pp+(-1)^n\frac{\q}{2}\right)\right\}
\times \nonumber\\
 \frac {t_{L'L}^*(p\cm_0,|\pp\cm|;0,|\pp^*|; s)}{(\sqrt
s/2-\ep+i\epsilon)^2-p_0^2}
g_l \left(p_0+(-1)^n\frac{\omega}{2},\pp+(-1)^n\frac{\q}{2};K_{(0)}\right)
\nonumber
\end{eqnarray}
corresponds to the process when FSI is taken into account. Here
$g_l$ denotes the radial part of the deuteron vertex function $\Gamma^M$. The part
${\rm Sp}\left\{\ldots\right\}$ has been calculated using the
algebra manipulation package MAPLE. The three-dimensional integration over $p_0\cm$, $|\pp\cm|$ and
$\cos\thetap\cm$ has been performed numerically using the
programming language FORTRAN.

%
% \cite{Breuker:1985sz},\cite{Boden:1992jz}
{\small
\begin{table}[ht]
\begin{center}
\begin{tabular}{|ll|c|c|c|c|c|c|c|c|}
\hline
            &         &  \boa   &  \bob   &  \boc   &  \bod   &  \boe  \\
\hline
\ee, GeV
            &         &  1.464  &  1.569  &   1.2   &    1.2  &   1.2  \\
\hline
\eepr, GeV
            & {min} &  1.175  &  1.118  &  0.895  &  0.895  &  0.895 \\
            & {max} &         &         &  0.800  &  0.800  &  0.800 \\
\hline
\thepr, ${}^{\circ}$
            &         &    21   &    21   &  20.15  &  20.15  &  20.15 \\
\hline
$\pp_n$, GeV/$c$
            & {min} &  0.314  &  0.500  &  0.126  &  0.197  &  0.197 \\
            & {max} &  0.660  &  0.773  &  0.564  &  0.423  &  0.488 \\
\hline
\thn, ${}^{\circ}$
            & {min} &  60.53  &  74.60  & 142.32  & 155.72  & 165.36 \\
            & {max} &  62.49  &  63.52  &  93.96  & 136.09  & 112.86 \\
\hline
\thqe, ${}^{\circ}$
            & {min} &  61.94  &  45.57 &   59.56  &  51.52  &  51.25 \\
            & {max} &  37.39  &  29.49 &   25.57  &  25.57  &  25.57 \\
\hline
$\pp_p$, GeV/$c$
            & {min} &  0.466  &  0.681  &  0.525  &  0.620  &  0.622 \\
            & {max} &  0.664  &  0.791  &  0.834  &  0.929  &  0.889 \\
\hline
\thp, ${}^{\circ}$
            & {min} &  35.82  &  45.12  &   8.42  &   7.52  &   4.47 \\
            & {max} &  61.68  &  60.90  &  42.40  &  18.41  &  30.40 \\
\hline
\thpe, ${}^{\circ}$
            & {min} &  97.77  &  90.68  &  68.00  &  44.00  &  56.00 \\
            & {max} &  99.08  &  90.39  &         &         &        \\
\hline
\sqs, GeV
            & {min} &  1.9675 &  2.1375 &  1.98   &  2.04   &  2.04  \\
            & {max} &  2.2125 &  2.3325 &  2.28   &  2.28   &  2.28  \\
\hline
\sqsmm, GeV
            & {min} &  0.090  &  0.260  &  0.101  &  0.161  &  0.161 \\
            & {max} &  0.335  &  0.455  &  0.401  &  0.401  &  0.401 \\
\hline
$Q^2$, (GeV/$c$)$^2$
            & {min} &  0.257  &  0.255  &  0.154  &  0.145  &  0.145 \\
            & {max} &  0.206  &  0.209  &  0.106  &  0.106  &  0.106 \\
\hline
$\omega$, GeV
            & {min} &  0.162  &  0.348  &  0.148  &  0.210  &  0.210 \\
            & {max} &  0.422  &  0.568  &  0.476  &  0.476  &  0.476 \\
\hline
$|\q|$, GeV/$c$
            & {min} &  0.532  &  0.613  &  0.420  &  0.435  &  0.435 \\
            & {max} &  0.620  &  0.729  &  0.577  &  0.577  &  0.577 \\
\hline
\end{tabular}
\end{center}
\caption{ {\small Kinematic conditions considered in the paper.
Here all quantities are given in LS. In addition to those which are
defined in the text, they are: angle $\theta_{qe}$ between the
beam and the virtual photon; neutron momentum $\pp_n$ and angle
$\theta_n$ between the neutron and the virtual photon
($\pp_p,\theta_p$ -- the same for the proton); $\theta_{pe}$
($\theta_{qe}$) -- the angle between the beam and the proton
(virtual photon).} }\label{tab:2}
\end{table}
}

\section{Results and discussion}\label{sec:5}

The calculations of the deuteron electrodisintegration within
PWA were considered in~\cite{Bondarenko:2010qv,Bondarenko:2005rz}.
As it was shown in~\cite{Shebeko,Rogochaya-ZLetters},
a contribution of FSI effects increases with increasing nucleon energies
or/and momentum transfer. The reaction near the
threshold was considered  in \cite{Rogochaya-ZLetters} (Sacley experiments,
see~\cite{Bussiere:1981mv,TurckChieze:1984fy}).

\begin{figure}[h]
\begin{minipage}{0.48\textwidth}
 \includegraphics[width=0.99\textwidth]{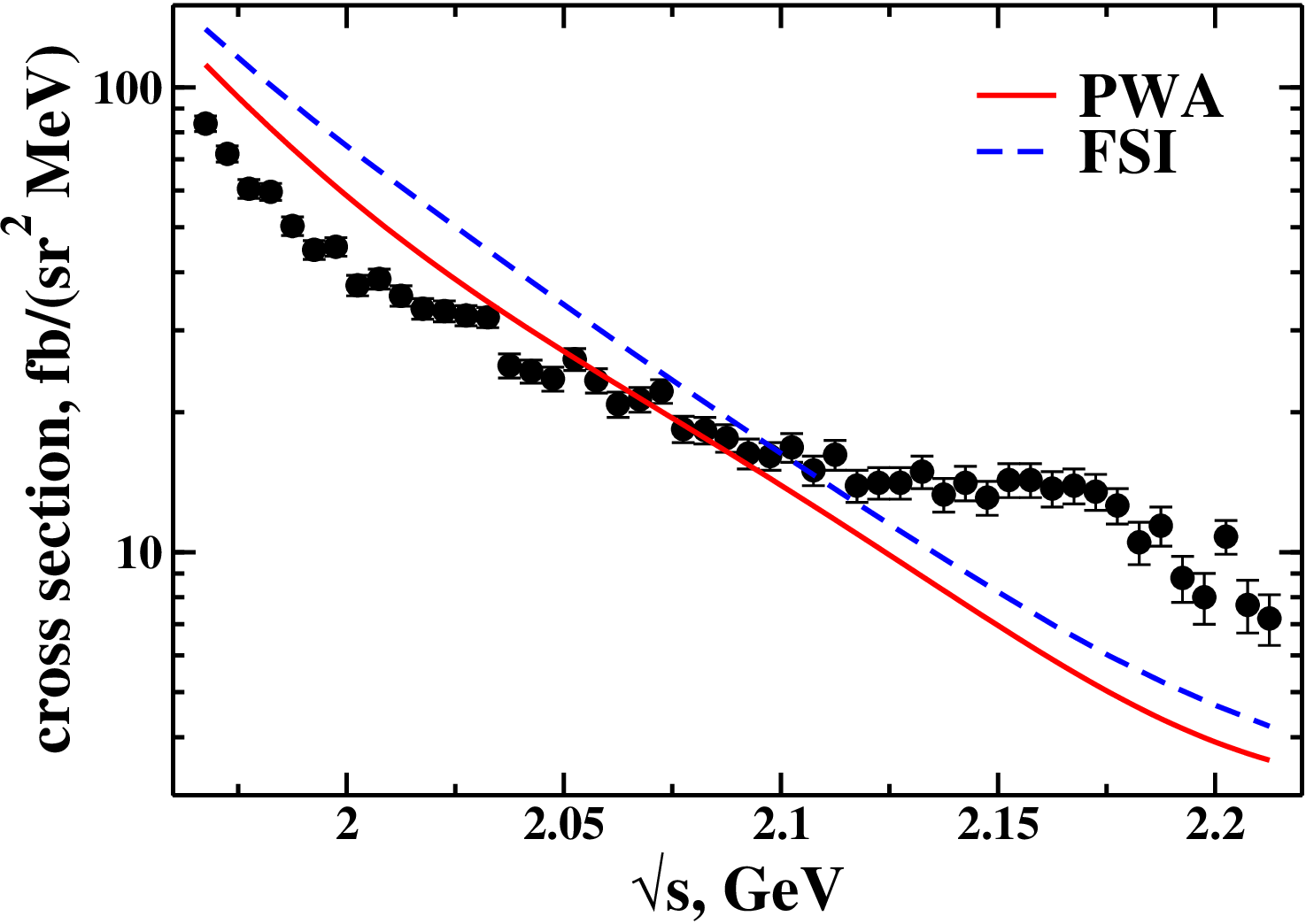}
\caption{
{\small Cross section (\ref{3cross_0}) depending on $\sqrt{s}$
calculated under kinematic conditions set I
of the Bonn experiment \protect\cite{Breuker:1985sz}.
%The notations are following: MY6 (PWA) (red solid line) -
%relativistic calculation in the plane-wave approximation with the
%MY6 potential \protect\cite{Bondarenko:2010qv}; MY6 (FSI) (blue dashed line)
%- relativistic calculation including FSI effects.
}}\label{fig:1}
\end{minipage}
\vskip -77mm
\hskip 82mm
\begin{minipage}{0.48\textwidth}
 \includegraphics[width=0.99\textwidth]{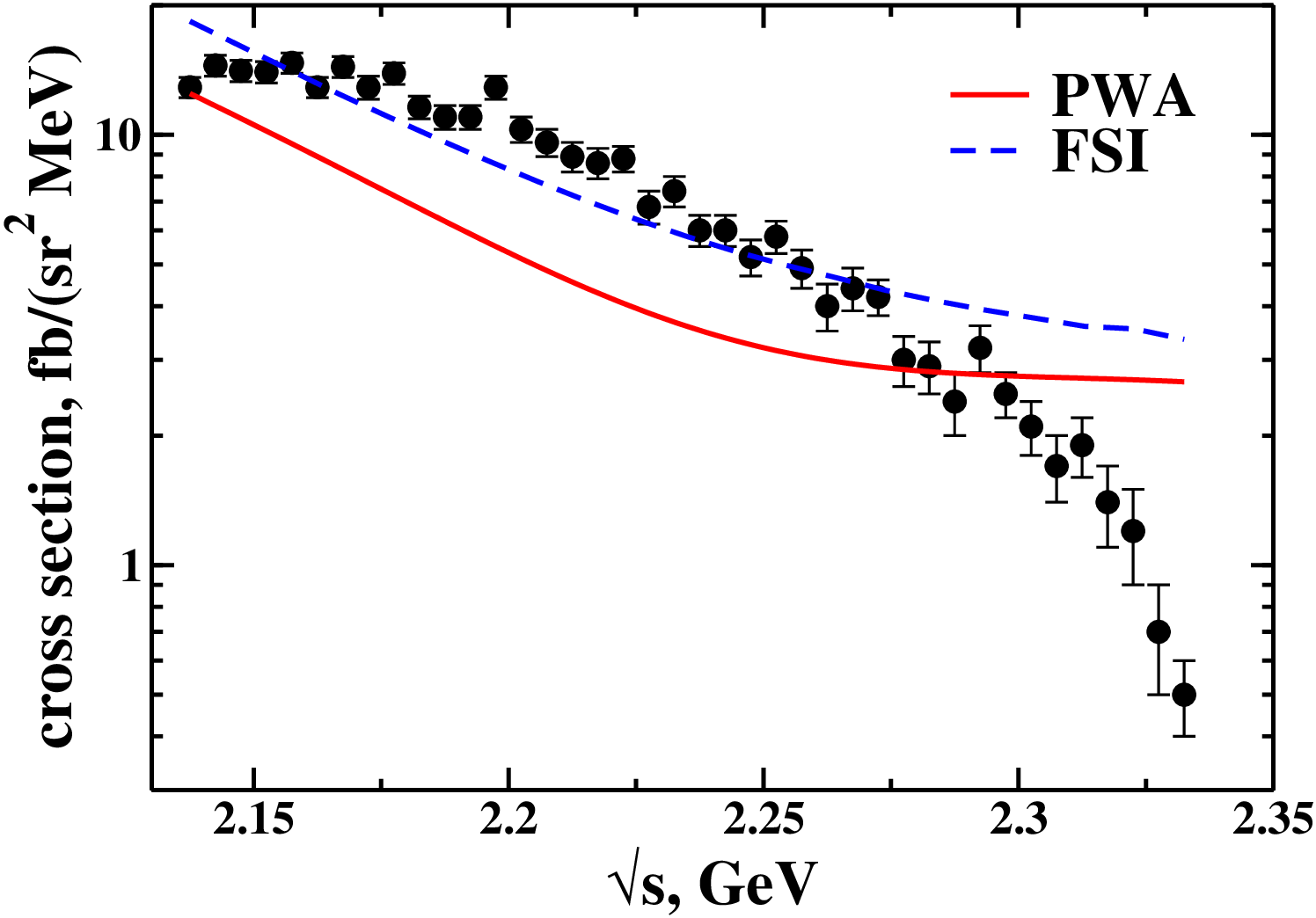}
\caption{
\small The same as in Fig.\ref{fig:1} but under kinematic
conditions set II of the Bonn experiment \protect\cite{Breuker:1985sz}.
}\label{fig:2}
\end{minipage}
\end{figure}

\begin{figure}[h]
\begin{minipage}{0.48\textwidth}
 \includegraphics[width=0.99\textwidth]{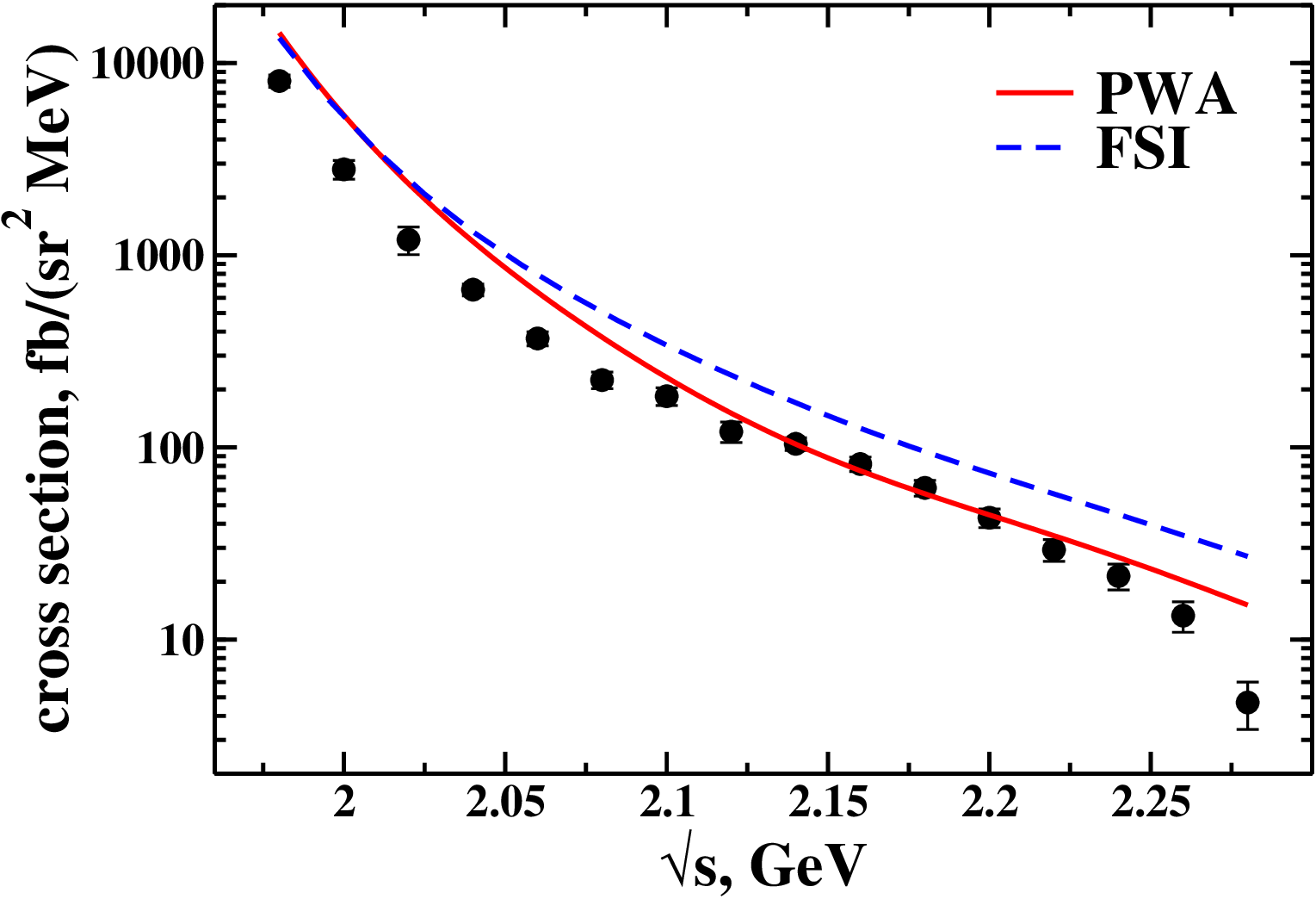}
\caption{
{\small Cross section (\ref{3cross_0}) depending $\sqrt{s}$
calculated under kinematic conditions set III
of the Bonn experiment \cite{Boden:1992jz}.}
}\label{fig:3}
\end{minipage}
\vskip -75mm
\hskip 82mm
\begin{minipage}{0.48\textwidth}
 \includegraphics[width=0.99\textwidth]{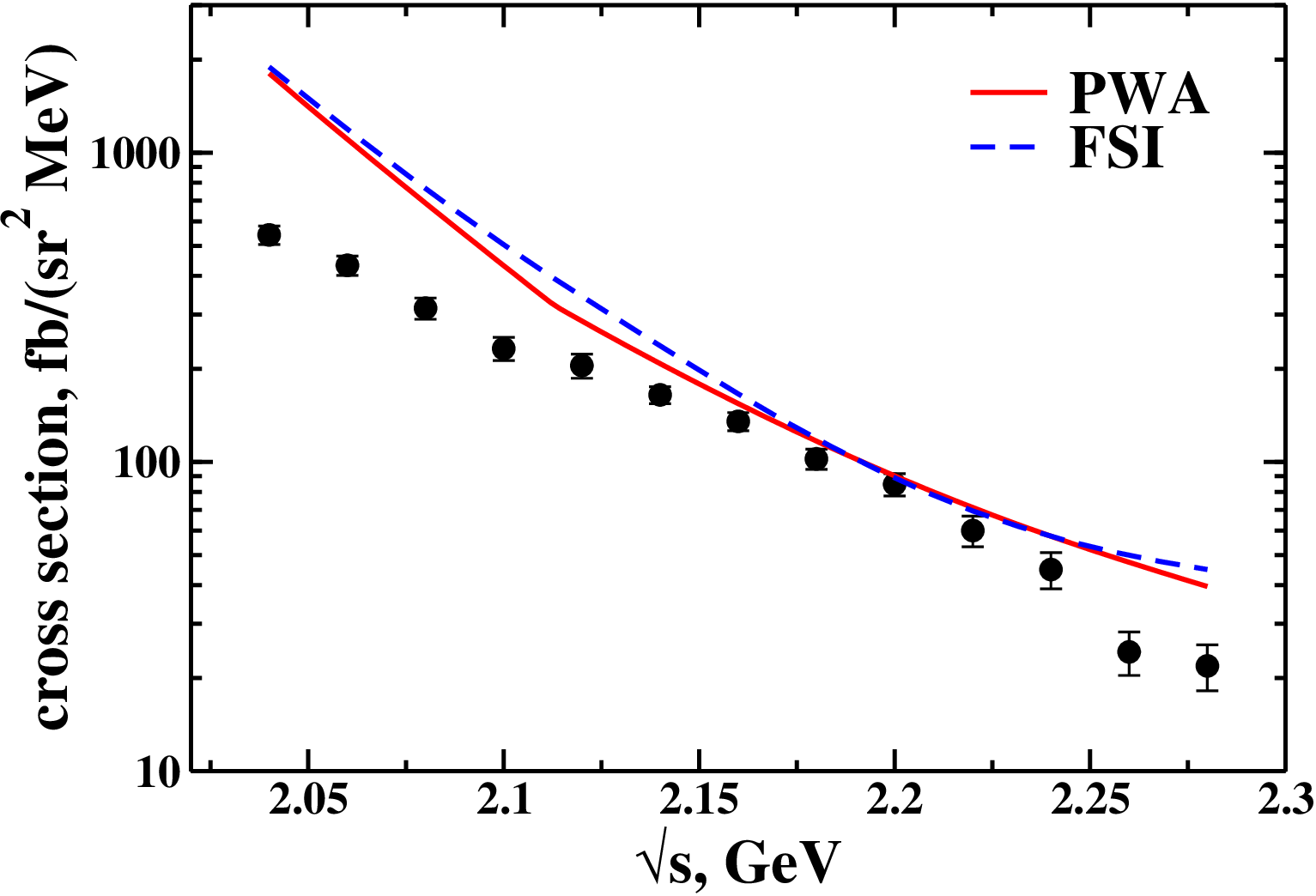}
\caption{
\small The same as in Fig.\ref{fig:1} but under kinematic
conditions set IV of the Bonn experiment \cite{Boden:1992jz}.
}\label{fig:4}
\end{minipage}
\end{figure}

\begin{center}
\begin{figure}[h]
\begin{minipage}{0.48\textwidth}
 \includegraphics[width=0.99\textwidth]{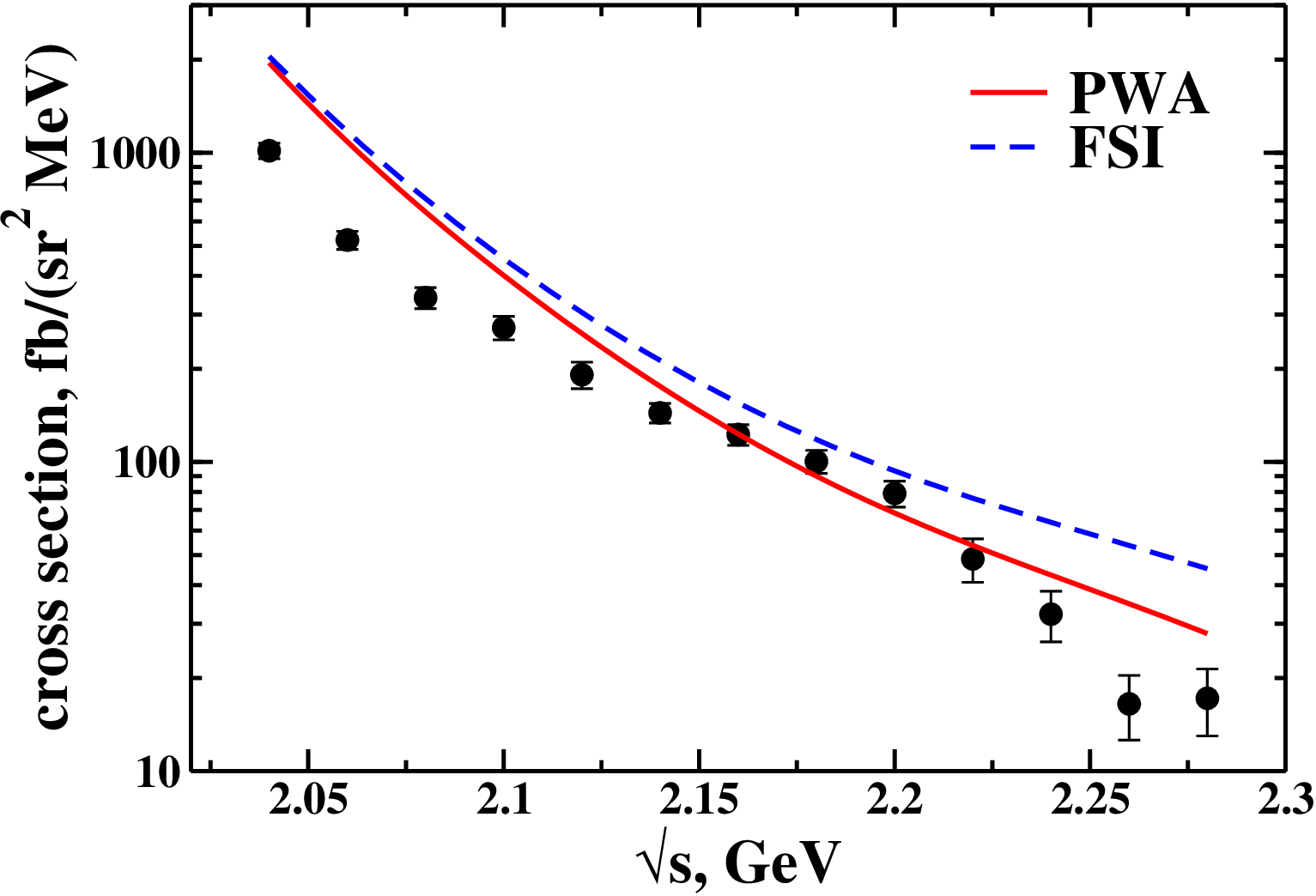}
\caption{
{\small Cross section (\ref{3cross_0}) depending on $\sqrt{s}$
calculated under kinematic conditions set V
of the Bonn experiment \cite{Boden:1992jz}.}
}\label{fig:5}
\end{minipage}
\end{figure}
\end{center}

In the paper the differential cross section (\ref{3cross_0}) has been
calculated under five kinematic conditions of the Bonn experiments
\cite{Breuker:1985sz,Boden:1992jz}
(described in Table \ref{tab:2}) and is shown in
Figs.\ref{fig:1}-\ref{fig:5}. The calculations have been performed
within the relativistic impulse approximation for two different
cases: when the outgoing nucleons are supposed to be free (PWA)
and when the final state interaction between the nucleons is taken
into account (FSI). The partial-wave states of the $np$ pair with
total angular momentum $J=0,1$ have been considered. The used
relativistic model consists of two parts: the separable potential MY6
\cite{Bondarenko:2010qv} for the bound (deuteron) and scattered
$^3S_1$-$^3D_1$ states and separable potentials of various ranks
\cite{Bondarenko:2008mm} - for all other partial-wave states
($^1S_0$, $^1P_1$, $^3P_0$, $^3P_1$).

In Figs.\ref{fig:1}-\ref{fig:5}, relativistic PWA (solid red line)
and FSI (dashed blue line) calculations are shown as a function of
$\sqrt{s}$.

As it is seen from Figs.\ref{fig:1}-\ref{fig:5}, the effect of FSI
increases the cross section for kinematic conditions
\cite{Breuker:1985sz,Boden:1992jz}.

In Fig.\ref{fig:2}, the calculation with FSI describes the experimental
data on $\sqrt{s}$ from 2.15 GeV to 2.3 GeV and differs
from experimental data starting from 2.3 GeV. Calculations of the cross
section shown in Figs.\ref{fig:1}-\ref{fig:2} require additional
investigations at high $s$ (for instance, the influence of non-nucleon
degrees of freedom).

In Figs.\ref{fig:3}-\ref{fig:5}, PWA and FSI calculations under the kinematic
conditions~\cite{Boden:1992jz} are shown. The effect of FSI is small
at the threshold but it becomes bigger with increasing $\sqrt{s}$.

The inelasticity effects are nonzero for kinematic conditions
of Bonn experiments~\cite{Breuker:1985sz,Boden:1992jz}.
However, they will be discussed in a separate paper.

%\newpage
\phantom{aaa}

\vskip 6mm

\newpage
\phantom{aaa}
%\vskip 1mm

\end{document}